\documentclass[twocolumn]{aastex61}

\newcommand\aastex{AAS\TeX}

\received{July 31, 2017}
\revised{September 11, 2017}
\accepted{September 11, 2017}
\submitjournal{ApJ}

\shorttitle{\aastex\ Collision Experiments: Ice Characterization }
\shortauthors{G{\"a}rtner et al.}

\begin{document}

\title{Micrometer-Sized Water Ice Particles for Planetary Science Experiments:\\
Influence of Surface Structure on Collisional Properties}

\email{sabrina.gaertner@stfc.ac.uk, helen.fraser@open.ac.uk}

\author[0000-0002-0786-5475]{S. G{\"a}rtner}
\affiliation{School of Physical Sciences, The Open University, 
	Walton Hall, Milton Keynes MK7 6AA, UK}

\author{B. Gundlach}
\affiliation{Institut f{\"u}r Geophysik und extraterrestrische Physik, 
	TU Braunschweig, Mendelssohnstr. 3, 38106 Braunschweig, Germany}

\author[0000-0003-0095-5731]{T. F. Headen}
\affiliation{ISIS Facility, STFC Rutherford Appleton Laboratory, 
	Harwell Oxford, Didcot OX11 0QX, UK}

\author{J. Ratte}
\affiliation{Institut f{\"u}r Geophysik und extraterrestrische Physik, 
	TU Braunschweig, Mendelssohnstr. 3, 38106 Braunschweig, Germany}

\author{J. Oesert }
\affiliation{Zoologisches Institut, Christian-Albrechts-Universit{\"a}t zu Kiel, 
	Am Botanischen Garten 1-9, 24118 Kiel, Germany}

\author{S. N. Gorb}	
\affiliation{Zoologisches Institut, Christian-Albrechts-Universit{\"a}t zu Kiel, 
	Am Botanischen Garten 1-9, 24118 Kiel, Germany}

\author[0000-0003-3538-5572]{T. G. A. Youngs}
\affiliation{ISIS Facility, STFC Rutherford Appleton Laboratory, 
	Harwell Oxford, Didcot OX11 0QX, UK}

\author[0000-0002-4557-1929]{D. T. Bowron}
\affiliation{ISIS Facility, STFC Rutherford Appleton Laboratory, 
	Harwell Oxford, Didcot OX11 0QX, UK}

\author[0000-0003-1531-737X]{J. Blum}
\affiliation{Institut f{\"u}r Geophysik und extraterrestrische Physik, 
	TU Braunschweig, Mendelssohnstr. 3, 38106 Braunschweig, Germany}

\author[0000-0003-0972-1595]{H. J. Fraser}
\affiliation{School of Physical Sciences, The Open University, 
	Walton Hall, Milton Keynes MK7 6AA, UK}





\begin{abstract}

Models and observations suggest that ice-particle aggregation at and 
beyond the snowline dominates the earliest stages of planet-formation, 
which therefore is subject to many laboratory studies. 
However, the pressure-temperature gradients in proto-planetary disks mean 
that the ices are constantly processed, 
undergoing phase changes between different solid phases and the gas phase.
Open questions remain as to whether the properties of the icy particles 
themselves dictate collision outcomes and therefore how effectively 
collision experiments reproduce conditions in protoplanetary environments.
Previous experiments often yielded apparently contradictory results 
on collision outcomes, only agreeing in a 
temperature dependence setting in above $\approx$~210~K.

By exploiting the unique capabilities of the NIMROD neutron scattering 
instrument, we characterized the bulk and surface 
structure of icy particles used in collision experiments, 
and studied how these structures alter as a function of temperature 
at a constant pressure of around 30~mbar. 
Our icy grains, formed under liquid nitrogen, 
undergo changes in the crystalline ice-phase, sublimation, 
sintering and surface pre-melting as they are heated from 103 to 247~K.
An increase in the thickness of the diffuse surface layer from  
$\approx10$ to $\approx30~\mathrm{\AA}$ ($\approx2.5$ to $12$ bilayers) 
proves increased molecular mobility at temperatures above $\approx$~210~K.
As none of the other changes tie-in  
with the temperature trends in collisional outcomes, 
we conclude that the surface pre-melting phenomenon plays a key role 
in collision experiments at these temperatures.
Consequently, the pressure-temperature environment, 
may have a larger influence on collision outcomes than previously thought.

\end{abstract}





\section{Introduction}\label{sec:introduction}

Dust aggregation is a key step in planet-formation 
\citep{testi_2014, garaud_2013, blum_2008, wada_2008}, 
enhanced by water ice at and beyond the snowline 
\citep{kataoka_2013, aumatell_2011, gundlach_2015}. 
But, we can only quantify collisional outcomes empirically, 
to learn how icy dust sticks under proto-planetary conditions 
\citep[e.g.][]{gundlach_2015, hill_2015a, bridges_1996, higa_1996}. 

Protoplanetary disk models indicate that icy particles are continually 
processed as particles traverse temperature and density gradients 
\citep{woitke_2016, woitke_2015, visser_2009}, 
resulting in repeated evaporation and re-formation of the water ice, 
which may be amorphous, crystalline or a mixture of both
\citep{sirono_2017, sirono_2011a, sirono_2011b, ros_2013}. 
Indeed, both types have been observed in accretion disks
\citep{boogert_2015, terada_2012, schegerer_2010}. 
Previous experiments 
\citep{wang_2005} 
showed that crystalline ice films absorb much less energy from  
impacts than amorphous ices, implying that collisional outcomes between 
proto-planetary disk particles could vary as a function of ice phase. 

However, all laboratory experiments necessarily have to use analogs 
rather than interstellar ice, 
and the ice particle formation mechanisms in the laboratory diverge 
from those in astronomical environments.
We therefore address the following outstanding questions:
\begin{itemize}
	\item Are the icy particles we are colliding in laboratory experiments 
	good analogs for protoplanetary disk environments? 
	\item Does the ice phase of our particles affect the collisional outcome? 
	\item	Does the surface structure play a dominant role?
\end{itemize} 

We have exploited neutron scattering and cryo-SEM 
(scanning electron microscopy) to characterize the ice 
particle analogues used in our laboratory collision experiments 
\citep{gundlach_2015} 
to ascertain whether phase changes in the bulk ice, 
and/or surface structural changes,
tie-in with the temperature-dependencies observed in collisional data, 
and whether the production method of the ice analogs 
influences the particle structure.

In combination, these data reveal 
which of the ice properties can affect collisional outcomes and 
to what extent these properties are altered by the collision environment. 
This information is essential to relate laboratory data 
back to planet-formation scenarios and to disentangle 
the seemingly contradictory results from laboratory collision experiments 
performed under different conditions, such that 
the most appropriate data can be employed in planet-forming models, 
and where necessary such models can be modified 
to account for the influence of ice physics on collision outcomes.


\section{Outstanding Challenges from Empirical Ice Collision Data}
\label{sec:problem}

Generally, planet formation requires aggregation of small particles 
to form bigger ones. 
However, particle sticking (a perquisite of models) is observed only 
in a small subset of collision experiments 
and then over a range of sticking probabilities 
\citep[$20-100~\%$;][]{deckers_2016, musiolik_2016, gundlach_2015, 
shimaki_2012b, bridges_1996, hatzes_1991}. 
Interestingly, all these studies were performed at relatively high pressures 
($1-10^3$~mbar), 
so that we cannot know whether the results would have been the same at 
lower pressures as expected beyond the snowline of proto-planetary disks 
\citep[$<1$~mbar;][]{cieza_2016}.
 
All laboratory experiments where sticking is observed, 
have in common that they involved micrometer-sized structures 
(small particles or layers of condensed water often referred to as ``frost''). 
Indeed, models predict that the particle size 
strongly influences the sticking probability during collisions: 
the sticking threshold velocity $v_\mathrm{stick}$ 
decreases with increasing particle radius $r$: 
$v_\mathrm{stick}\propto r^{-2/3}$ for 0.1 to 10~$\mu$m-sized particles 
\citep[see Figure 12 in][]{gundlach_2015}, 
and $v_\mathrm{stick}\propto r^{-1}$ for mm- to m-sized particles
\citep[see Figure 7.1 in][]{heisselmann_2015}.
However, micrometer-sized features on the surface of cm-sized particles 
(as induced by roughening or water condensation) are far less predictable and
several collisional studies on such particles did not observe any sticking
\citep{dilley_1996, higa_1996, mcdonald_1989, hatzes_1988}. 

In such collision experiments that do not lead to sticking, 
the coefficient of restitution, $\epsilon$, is extracted, 
which describes the loss of translational energy resulting from the collision,
and eventually feeds into models of planet-formation.
However, previous experiments
\citep{bridges_1984, hatzes_1988, mcdonald_1989, hatzes_1991, 
supulver_1995, dilley_1996, higa_1996, bridges_1996, higa_1998, 
heisselmann_2010, shimaki_2012a, shimaki_2012b, hill_2015a, 
gundlach_2015, deckers_2016, musiolik_2016} 
disagree on whether, and how, $\epsilon$ varies as a function of temperature, pressure, velocity, size, and shape.
The question is, why is this?
We hypothesize that two key factors play a role: 
first the method and prevailing conditions 
under which the icy particles are formed, 
and second the prevailing conditions 
under which the particle collisions are investigated.

The two key environmental parameters 
are pressure $P$ and temperature $T$. 
However, these two parameters are not usually varied systematically, 
resulting in contradictory experimental results 
and making it difficult to ascertain 
exactly which empirical data are most relevant to planet-forming models. 
From the few cases where $T$ has been varied at constant $P$ 
\citep{bridges_1984, mcdonald_1989, higa_1996, 
heisselmann_2010, gundlach_2015, hill_2015a}, 
two clear trends are evident; the collisional outcomes are 
temperature-independent below $T\approx210$~K 
\citep[e.g.][]{gundlach_2015, heisselmann_2010, hill_2015a} 
and become temperature-dependent above $T\approx210$~K 
\citep[e.g.][]{gundlach_2015}, 
where the coefficient of restitution decreases and the 
threshold velocity to particle sticking increases, as temperature increases. 

The ice projectiles in these collision experiments  
have been formed under various conditions, 
but always from the liquid phase. 
While freezing of liquid water in a kitchen freezer or under liquid nitrogen 
is expected to yield some form of crystalline ice, 
the ice structure on a molecular scale will depend on the freezing rate 
as well as on the conditions (and duration) under which the ice was 
processed and/or stored between initial freezing and eventual collision.
This ``thermal history'' again has not always been varied systematically 
and not even always been fully described.

Micrometer-sized particles can be created 
by shattering larger bodies of ice prepared in a freezer 
\citep{deckers_2016}, 
or by rapid freezing of water droplets e.g.\ on cold surfaces 
\citep{musiolik_2016} 
or by introducing them into a cold gaseous or liquid environments
\citep{gundlach_2015, shimaki_2012b}, 
However, without further characterization, 
we cannot know whether, and how, 
the production alters the particle structure on all length-scales, 
exactly which form of crystalline ice is produced, and to what extend 
the $P$-$T$ conditions of the collision environment 
influence the collisional outcomes.

As no definitive particle characterizations have been made to date, 
ice phase and micro-scale structure are alluded to 
in icy particle collision experiments and their influence on collision outcomes 
remains a contentious issue in the literature, which we address in this work. 

By exploiting the unique capabilities of the NIMROD 
\citep{bowron_2010} 
neutron scattering instrument, 
we characterized the bulk and surface structure of the icy particles.
NIMROD can simultaneously observe a wide range of length scales 
from the mesoscale ($\approx60$~nm) down to the intra-molecular level, 
which means it is possible to concurrently establish the phase, 
molecular structure and surface properties of icy materials 
\citep{mitterdorfer_2014, hill_2016}.


\section{Experimental Method}\label{sec:experimental}

The particles for this characterization study were produced 
as described in detail in
\citet{gundlach_2011, jost_2013, gundlach_2015}.
Briefly, liquid $\mathrm{D_2O}$ was dispersed by an aspirator 
and sprayed into liquid nitrogen, 
accumulating sample material for $>1$~hour. 
The particles were then funneled into the pre-cooled (77~K) 
sample container, which was closed, mounted in the neutron beam, 
and passively cooled at a constant $P$ (30~mbar He).

We studied samples with two different mean particle radii, 
($(0.71\pm0.31)~\mu$m and $(1.45\pm0.65)~\mu$m), 
where uncertainties give the particle size distribution's 
FWHM (full width at half maximum)
\citep{gundlach_2011}. 
Two independent experiments were conducted at each particle radius.

Advantage was taken of the neutron scattering properties of 
D$_2$O compared to H$_2$O 
\citep{sears_1992}, 
and care was taken to minimize sample contamination with 
H$_2$O during sample preparation and loading, 
by maintaining an N$_2$-purged environment, 
retaining both the container and sample below 100~K. 

Neutron scattering data was collected over 30~min isothermal periods, 
at 103, 164, 184, 206, 226, and 247~K. 
The initial data reduction and calibration was done using GudrunN software 
\citep{soper_2011, soper_2013}, 
according to standard neutron scattering data processing. 
The raw data from our neutron scattering experiments were 
merged for all detectors, corrected for instrument effects, 
and normalized on a per atom basis.
Examples of the resulting background corrected neutron diffraction patterns 
are shown in Figure~\ref{fig:raw_data}(a) for one of the four experiments.
The small angle neutron scattering ($Q\leq0.1~\mathrm{\AA}^{-1}$), 
probing surface structures, 
was observed concurrently with the high-$Q$ region ($Q\geq1~\mathrm{\AA}^{-1}$), 
probing the bulk ice phase (intra-molecular distances). 

To give an impression of the particle structures on larger scales 
($0.1-100~\mu$m), we compared the neutron scattering results to images 
from complementary cryo-SEM experiments
(\citet{cryo_SEM}; see also \citet{jost_2013}), 
using H$_2$O-particles prepared the same way as previously described, 
but necessarily held at lower pressures ($P=10^{-3}$~mbar).


\section{Results}\label{sec:results}

   \begin{figure*}[]
   \includegraphics[width=\textwidth]{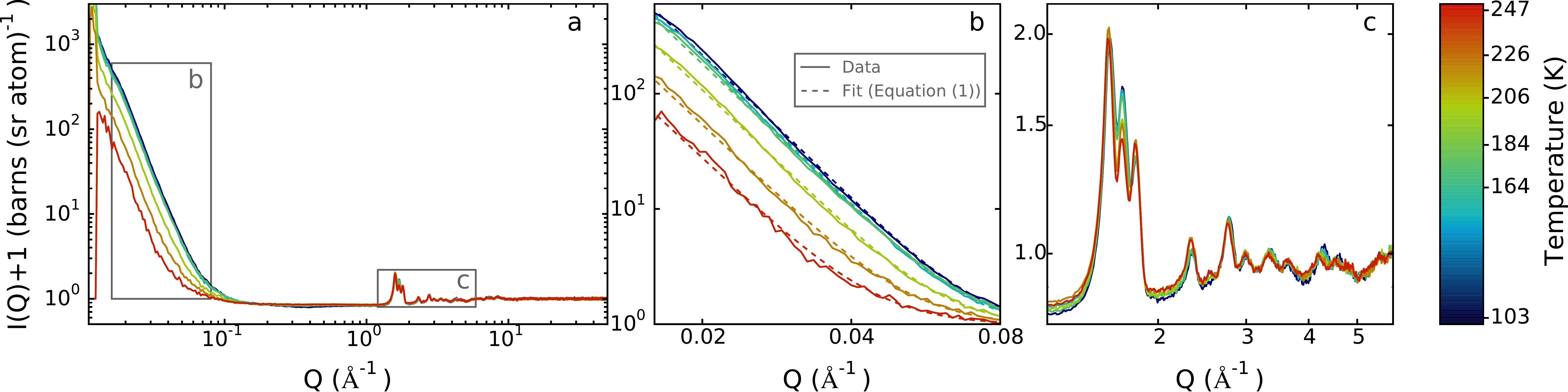}
      \caption{
		(a) NIMROD spectra ($0.71~\mu$m icy particles), 
		showing the neutron scattering signal $I(Q)$ (barns~(sr~atom)$^{-1}$) 
		as a function of the momentum transfer $Q$, 
		which is inversely proportional to the length scale.
		Sequential plots show the temperature evolution in six isothermal 
		steps between $103$ and $247$~K (color bar). 
		Expanded views: (b) low-$Q$ 
		(dashed lines show fit as per Equation~\ref{eq:diffuse_interface_fit}, 
		Section~\ref{sec:surface})
		and (c) high-$Q$ regions. 
		For clarity, error bars have been omitted 
		(average uncertainty: 6~\% of $I(Q)$).
		 }
         \label{fig:raw_data}
   \end{figure*}

Across the four neutron scattering experiments, 
no clear differences were seen between scattering from different icy 
particle sizes, nor in repeated experiments on particles of the same size. 
However, clear changes with increasing sample temperature are evident 
(Figure~\ref{fig:raw_data}), 
indicating temperature induced modifications in the icy particles, 
both in the bulk ice phase and the particle surfaces, 
which will be addressed in detail in the following. 

\subsection{Ice Phase}\label{sec:phase}

   \begin{figure}[bh]
   \includegraphics[width=\columnwidth]{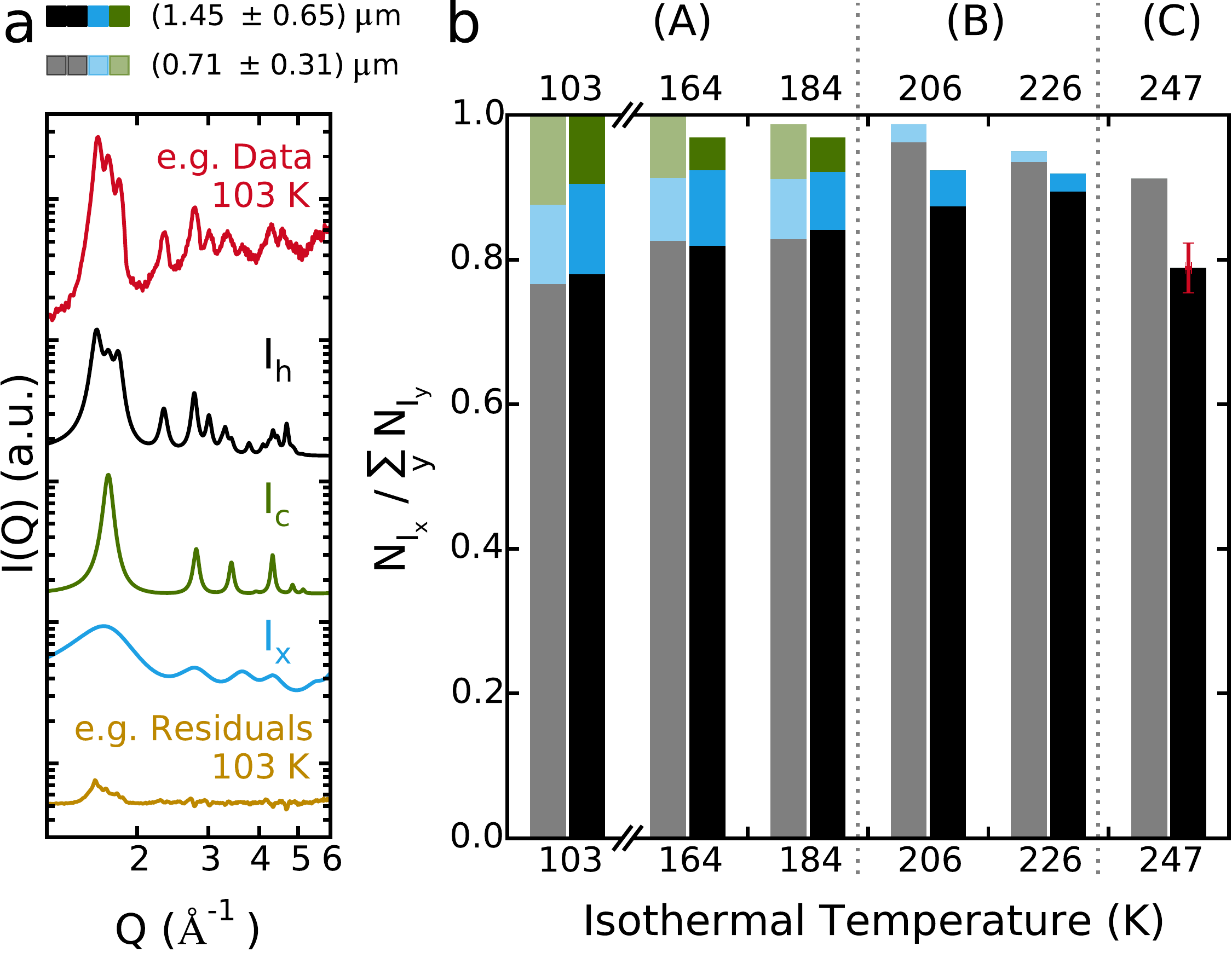}
   \caption{
		(a) original high-$Q$ data for the 103~K, $0.71~\mu$m particles (red), 
		together with diffraction patterns of the three ice phase components 
		$I_h$ (black), $I_c$ (green), and $I_{x}$ (blue). 
		Each high-$Q$ pattern was deconstructed into a sum of these 
		components as described in the text. 
		Also shown (yellow) are the residuals 
		when only $I_h$ and $I_c$ are fitted.
		(b) outcome of this analysis across our entire data set. 
		Cumulative bars represent the fractions of each ice-phase 
		(colors as in (a)) required to reproduce the 
		high-$Q$ data at each isothermal temperature, 
		averaged across two experimental runs, 
		and separated by particle radius (see legend). 
		Each cumulative bar is normalized to the total amount of ice 
		determined in each sample at the initial temperature point, 103~K. 
		The typical uncertainty, propagated from the individual fits, 
		is shown on the final $I_h$ bar; 
		for clarity other error bars have been omitted. 
		Dotted lines indicate the critical isothermal steps at which 
		drastic changes in the ice phase were observed, 
		breaking the data into three distinct ice-phase regimes; 
		$I_h+I_c+I_{x}$; $I_h+I_{x}$; $I_h$. 
		}
              \label{fig:ice_phase}%
    \end{figure}

The high-$Q$ region (Figure~\ref{fig:raw_data}(c)) shows peaks indicative 
of a crystalline-dominated ice structure; 
the most prominent feature is the triplet of Bragg peaks 
in exactly the positions expected for hexagonal ice, 
$I_h$, (1.59, 1.70, and 1.80~$\mathrm{\AA}^{-1}$ 
\citep{petrenko_1999}). 
There are some subtle but significant modifications to this triplet as a 
function of temperature; the peak at 1.70~$\mathrm{\AA}^{-1}$ actually 
corresponds to overlapping features from $I_h$ and $I_c$ 
(cubic crystalline ice) and loses intensity above 180~K. 
This is also reflected in changes to the smaller diffraction peaks 
at higher $Q$ and indicates that $I_c$ is lost.
There are two possible pathways for this loss; 
either transformation to the more stable $I_h$ or sublimation. 
Any increase in the amount of $I_h$ would result in an 
increase in the intensities of the other $I_h$ Bragg peaks, 
which appear unaltered.

To quantify these changes, we deconstructed each high-$Q$ 
diffraction pattern between $Q$ values of $1-6~\mathrm{\AA}^{-1}$
into a sum of features representing $I_h$ and $I_c$,
based on crystallographic calculations of the diffraction patterns 
for the respective idealized pure ice phases. 
The residuals of this analysis showed a very broad diffraction peak, 
whose shape and position closely matched that expected for 
amorphous ice (Figure~\ref{fig:ice_phase}(a) blue and yellow curves), 
so a third component was added to the deconstruction;
$I_{x}$, denoting inter-domain bulk material of no long-range order, 
which will be discussed in more detail in 
Section~\ref{sec:ice_phase_discussion}.

Within fitting uncertainties, this deconstruction showed that the 
temperature dependence of the ice structure is independent of 
particle radius, i.e.\ a bulk effect. 
Initially, the icy particles exhibit stacking disorder, dominated by $I_h$. 
They comprise areas of both $I_h$ and $I_c$ 
as well as inter-domain amorphous structures, $I_{x}$.
 
As $T$ increases, the normalized fractions of all three components change 
and the normalized fractions no longer sum to 1.
A measurable fraction of $I_c$ is retained until 184~K 
(Figure~\ref{fig:ice_phase}(b) regime (A)), 
but has essentially disappeared by 206~K. 
A fraction of $I_{x}$ persists until 226~K
(Figure~\ref{fig:ice_phase}(b) regime (B)); 
at 247~K the data is best fitted by $I_h$ only.

\subsection{Surface}\label{sec:surface}

  \begin{figure}
   \includegraphics[width=\columnwidth]{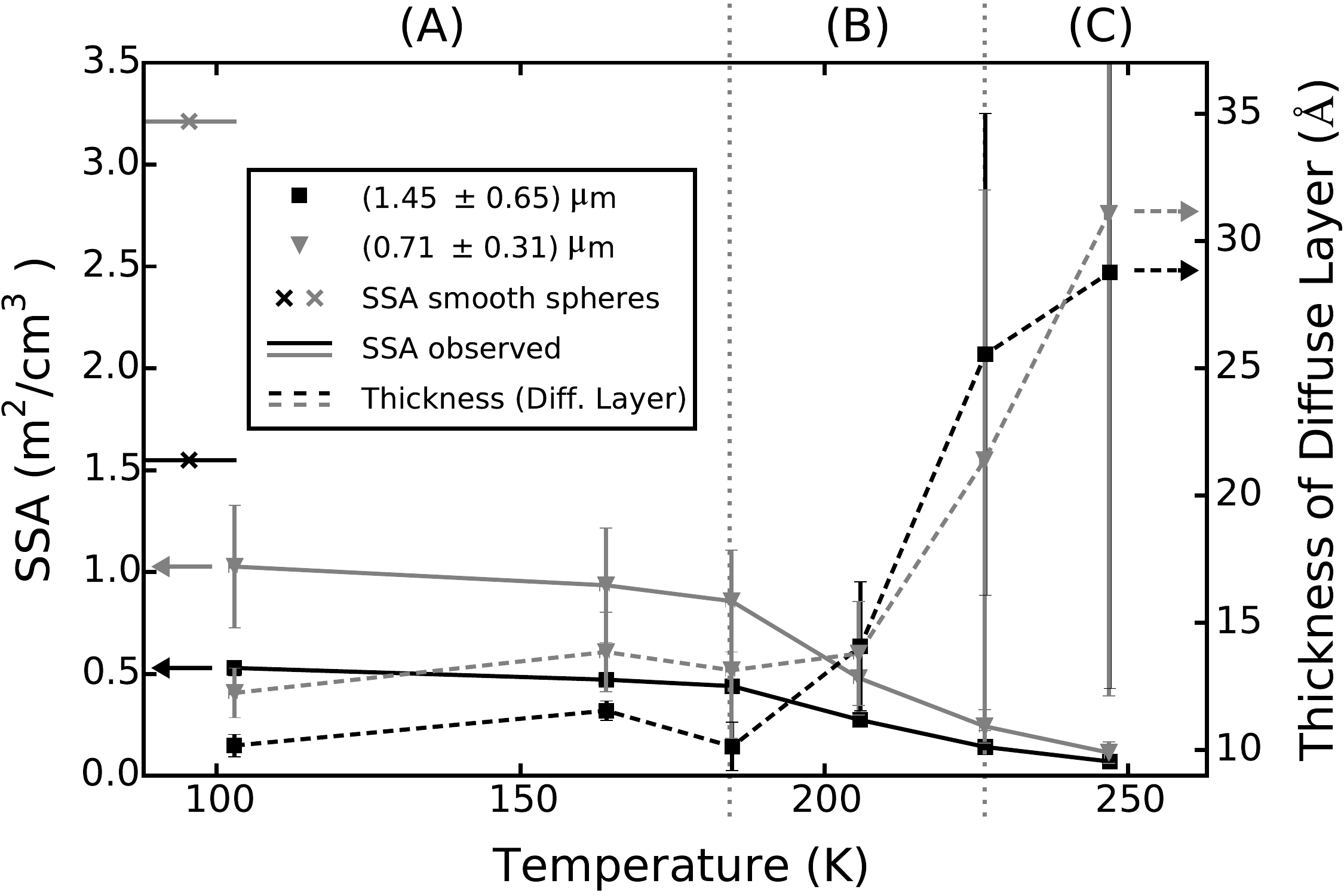}
      \caption{
		SSA and $t$, as a function of temperature, as extracted from the 
		low-$Q$ NIMROD data using Equation~\ref{eq:diffuse_interface_fit}. 
		For clarity, the results have been averaged per particle size 
		across experimental runs, and consecutive temperature points 
		joined with a solid line (SSA, left-hand axis) 
		or dashed line ($t$, right-hand axis). 
		Error bars represent the standard deviation of the mean.
		}
         \label{fig:SSA+thickness}
   \end{figure}

Returning to Figure~\ref{fig:raw_data}(b), 
the low-$Q$ data show much more drastic changes 
than the high-$Q$ Bragg peaks.
There is little, if any, obvious change in $I(Q)$ between 103 and 184~K, 
but as $T$ increases beyond 184~K, $I(Q)$ diminishes rapidly. 
This indicates a substantial change in the particles' surface structure, 
the onset of which coincides with the temperature regime at which 
$I_c$ is no-longer measurable in the bulk-ice.

At all temperatures the low-$Q$ slopes approximately follow the expected 
Porod power-law 
\citep{feigin_1987}, 
indicative of compact, granular material. 
This is expected from a sample of non-porous, 
spherical icy particles and thus confirms that formation in liquid Nitrogen 
does not alter the internal particle structures 
on length scales of tens of nanometers.

There are well established methods to extract quantitative information 
on the surface structures from these data.
Porod analysis \citep{sinha_1988} yields a Porod exponent $\beta$, 
which can be related to the ice surface roughness 
\citep[e.g.][]{mitterdorfer_2014, hill_2016}. 
For our icy particles, $\beta$ values ranged from 4.1 to 6, 
increasing non-linearly with increasing temperature. 
Values of $\beta>4$ indicate that no surface roughness on nm length scales 
is introduced by the freezing process, 
but that the surfaces are diffuse, i.e.\ showing a density gradient 
\citep[e.g.][]{strey_1991, su_1998}. 
However, in the case of diffuse interfaces the particle surface density cannot 
be validly modeled by a step function (as in the Porod analysis) but 
is best described by convoluting a Gaussian with said step function. 
The width of this Gaussian indicates the 
thickness of the diffuse interface, $t$. 
The resulting fit-function for the background corrected low-$Q$ data is 
\citep{strey_1991}: 
	\begin{equation}
	I(Q) = 2 \pi (\Delta \rho)^2\, \mathrm{SSA}\, Q^{-4} e^{-Q^2 t^2} ,
	\label{eq:diffuse_interface_fit}
	\end{equation}
where SSA is the specific surface area, and 
$\Delta \rho=5.995\times10^{-6} \mathrm{\AA}^{-1}$ is the 
scattering length density difference. 
Under our specific experimental conditions, 
this is the scattering length density of D$_2$O, 
since no other material is present that has not already been 
corrected for by the calibration scans.
The resulting fits ideally reproduce the experimental data over the entire 
low-$Q$ range, (Figure~\ref{fig:raw_data}(b): dashed lines).

A major advantage of this model is that it concurrently gives values of 
$t$ and SSA, (Figure~\ref{fig:SSA+thickness}). 
At all temperatures the SSA values are below those calculated for 
samples of smooth spherical particles with the given size distributions 
(Figure~\ref{fig:SSA+thickness} left-hand axis: 
light/dark $-\!\!\!-\mkern-18mu\times\!\!-$), 
which will be discussed in detail in Section~\ref{sec:surface_discussion}.
However, as expected, the SSA of the 0.71~$\mu$m particles is 
always greater than that of the 1.45~$\mu$m particles.
Regardless of particle size, 
the SSA slightly decreases in the $103-184$~K range; 
the most drastic changes in SSA set in beyond 184~K, 
then this loss rate slows beyond 226~K. 

It is interesting to note that when the regimes (dotted vertical lines) from 
Figure~\ref{fig:ice_phase} are transposed to 
Figure~\ref{fig:SSA+thickness}, the key temperatures at which 
ice-phase-compositional changes occur correspond exactly with 
the distinctive changes in SSA. 

Whilst changes in $t$ almost mirror those in SSA, 
the most drastic changes occur above 206~K, 
thus not matching the temperature regimes observed for 
ice phase and SSA. 
The absolute values of $t$ are closely comparable 
between the two particle sizes, starting from around 10~$\mathrm{\AA}$ 
at 103~K, which represents roughly 2.5 $I_h$ bilayers, and increase 
(on average) by a factor of 3 with temperature.

From the SSA and $t$ results alone, 
we cannot distinguish whether the surface changes are caused by 
particle sintering or by sublimation. 
However, the images obtained from our complementary cryo-SEM study 
can answer this question.

\subsection{Sintering or Sublimation?}\label{sec:SEM}

    \begin{figure*}[ht]
   \includegraphics[width=\textwidth]{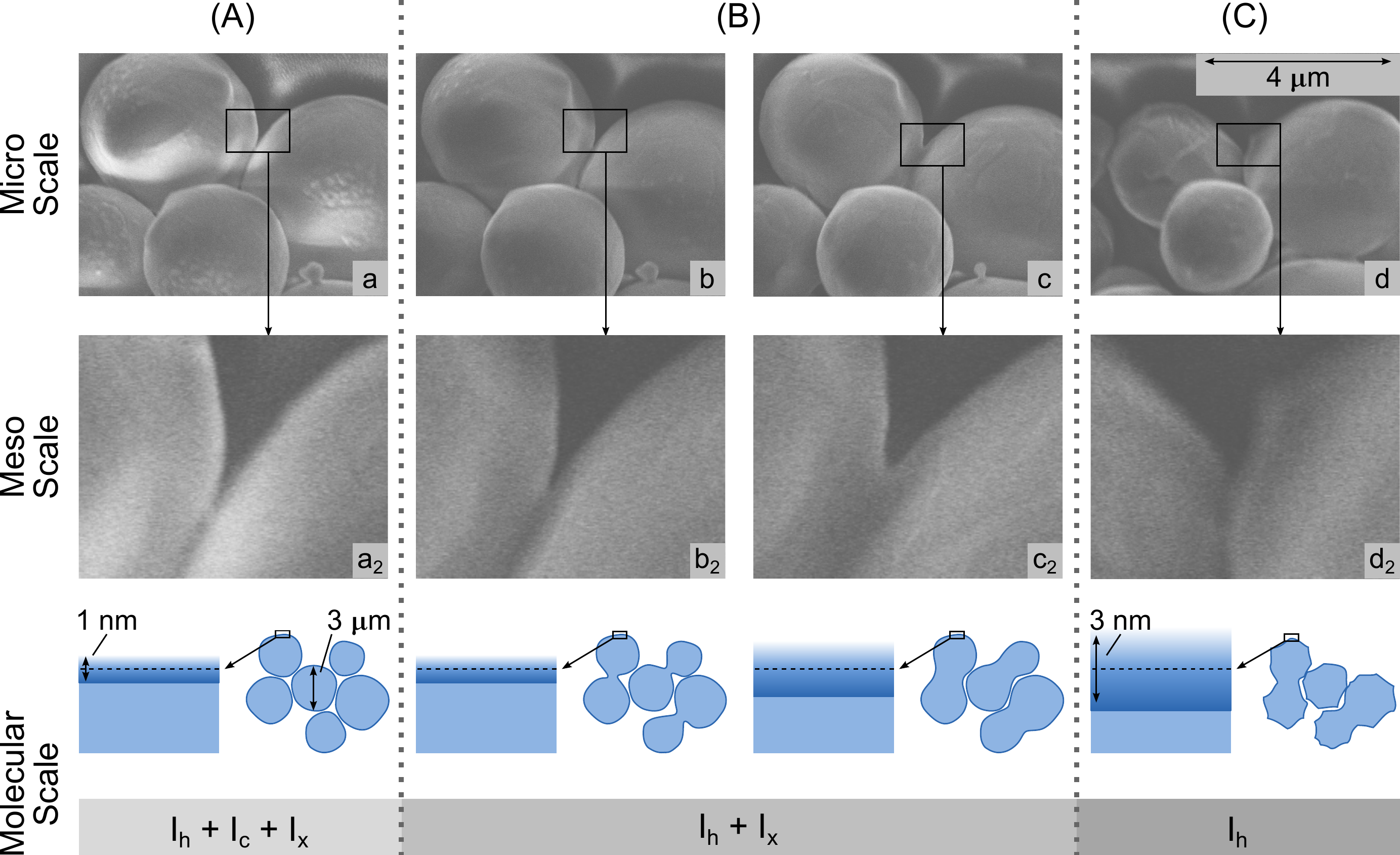}
   \caption{
		Schematic summarizing the temperature induced changes 
		observed in neutron scattering data from the icy particle samples 
		in the three regimes 
		(compare Figures~\ref{fig:ice_phase} and~\ref{fig:SSA+thickness}), 
		illustrating bulk and surface changes.
		Data are combined with cryo-SEM images, 
		indicating the visible changes to the particles on length scales 
		not accessible with NIMROD ($0.1-10~\mu$m).
		}
        \label{fig:ice_temp_change}%
     \end{figure*}

Figure~\ref{fig:ice_temp_change} joins up our findings on all length scales 
together with the SEM images. 
The vertical dotted lines indicate the same evolutionary stages in the 
ice-phase composition and SSA, 
as identified in Figures~\ref{fig:ice_phase} and~\ref{fig:SSA+thickness}.

The SEM images reveal that initially the particles are mostly, 
although not perfectly, spherical 
(Figure~\ref{fig:ice_temp_change}(a)--(c)). 
The icy particles are in contact with each other, but no sintering is evident. 
With increasing temperature, sintering is observed 
(Figure~\ref{fig:ice_temp_change}(b$_2$)), 
and as the temperature continues to rise, the sintering necks 
become more pronounced (Figure~\ref{fig:ice_temp_change}(c$_2$)).

Finally, by the highest temperatures, 
where the particles only comprise $I_h$ 
(i.e.\ beyond the second vertical dotted line in Figures~\ref{fig:ice_phase}, 
\ref{fig:SSA+thickness}, and~\ref{fig:ice_temp_change}), 
material seems to be lost from the narrow sinter-neck and 
the particles become faceted with straight edges and reduce in size 
(Figure~\ref{fig:ice_temp_change}(d$_2$)), 
while the smallest particles are lost.


\section{Discussion}\label{sec:discussion}

    \begin{figure*}[]
	\centering
   \includegraphics[width=\textwidth]{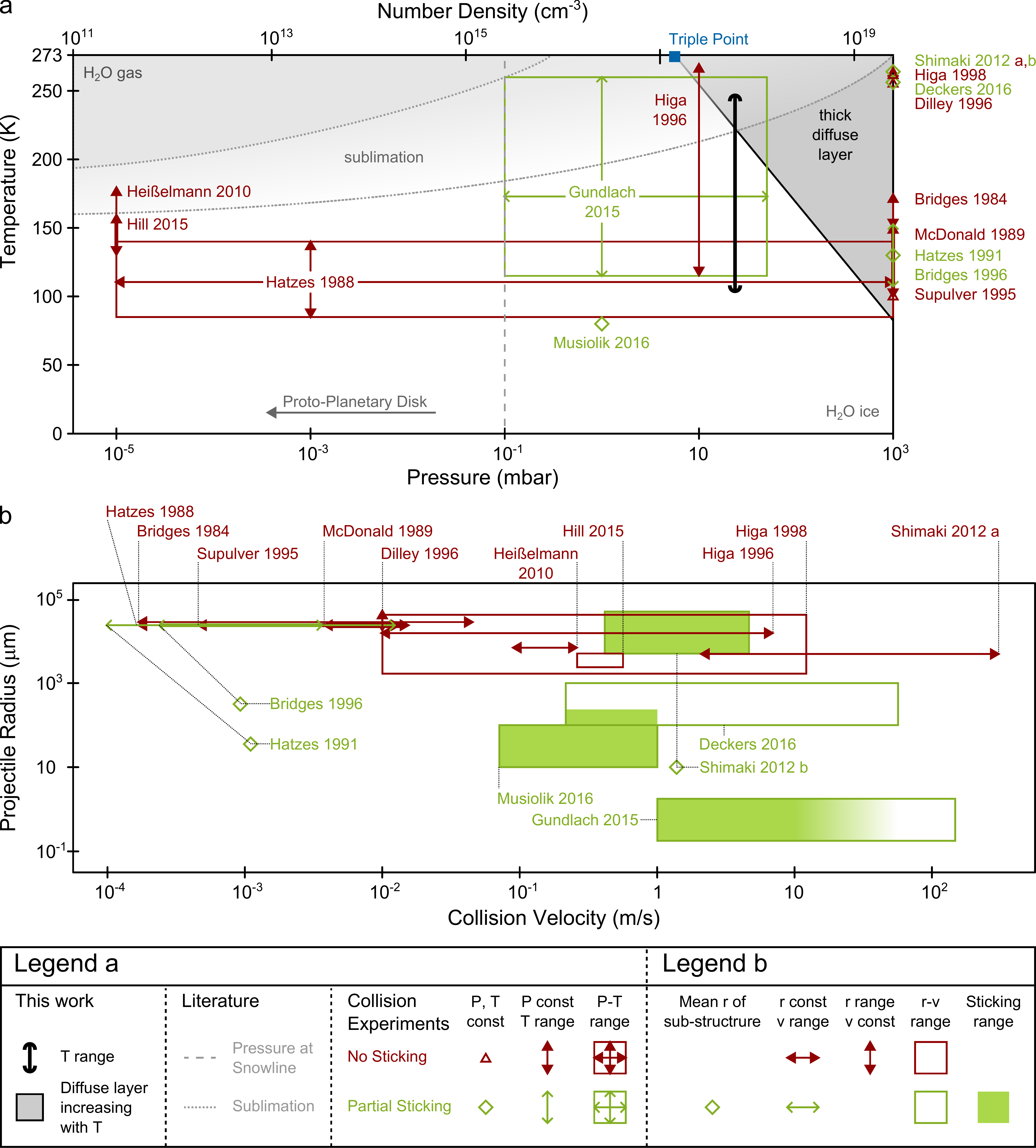}
   \caption{
		(a) schematic summarizing the $P$-$T$ conditions or ranges of 
		previous collision experiments (red and green symbols/lines), 
		together with those under which ice can exist (white area) 
		in the laboratory and in proto-planetary disks (PPDs).
		The region in which sublimation occurs 
		will depend on pressure and composition of the ambient gas 
		(more precisely on the partial H$_2$O pressure), 
		as depicted by the dotted lines.
		In PPDs the ice pressure-regime is limited by the snowline,  
		\citep[dashed line: snowline pressure for a typical PPD, see]
		[and references therein]{cieza_2016}.
		Thin diffuse surface layers were observed 
		across the whole temperature range of this work, 
		above a critical temperature these start to thicken as the ice is warmed 
		and strongly impact on collisional outcomes.
		This critical temperature will depend on the ambient pressure
		as (qualitatively) depicted by the dark gray area.
		(b) schematic summarizing the velocities, $v$, and radii, $r$, 
		used in the above collision experiments. 
		Partially sticking experiments are labeled within the schematic (green), 
		non-sticking experiments are labeled above the schematic (red). 
		The ranges across which partial sticking was observed are shaded 
		(green), the shading gradient on Gundlach~2015 
		indicates the observed T-variance. 
		For those studies where cm-sized projectiles 
		featured $\mu$m-sized structures 
		the mean size of these sub-structures is shown (diamonds) 
		and connected to the experiment ranges via dotted lines.
		}
        \label{fig:P_T_range}%
     \end{figure*}

We have characterized micrometer-sized icy particles identical to those 
used in laboratory collision experiments on planet-formation.
Our icy particles were produced under liquid Nitrogen and 
were not dissimilar in size to grains of crystallized water ice 
($\approx$~0.8~$\mu$m in size) 
that have been observed in the silhouette disk of a young star
\citep{terada_2012}.

Our characterization experiments cover almost the whole temperature range
exploited in laboratory collisional studies over the past three decades, from 80~K
\citep[e.g.][]{musiolik_2016, hatzes_1988} 
to 269~K
\citep[e.g.][]{shimaki_2012b, higa_1996},
performed to understand collisions in a variety of environments, 
like protoplanetary disks, cometary surfaces, and planetary rings.
Thus, our results provide crucial information towards the role of ice phase 
and surface structures in dictating collision outcomes in such environments.

\subsection{Does the ice phase affect the collisional outcome?}
\label{sec:ice_phase_discussion}

We find that our particles are initially stacking disordered,
as is expected when freezing water droplets in liquid Nitrogen 
\citep{malkin_2015, malkin_2012, kuhs_2012}. 
The particles comprise both low-pressure crystalline phases of ice, 
$I_c$ and $I_h$,
but the obtained diffraction patterns are best fitted 
when an amorphous ice phase is added.
Various amorphous ice candidates could be attributed to this third phase, 
e.g.\ HGW (hyper-quenched glassy water), 
LDA (low-density amorphous solid water), 
and HDA (high-density amorphous solid water), 
but the resolution of our diffraction data is not well suited 
to distinguish between them.
NIMROD was not designed as high resolution crystallography instrument, 
but rather to provide atomistically quantitative structural data for 
highly disordered and complex systems over a very wide $Q$-range.
Sophisticated models for the fitting of diffraction patterns from 
stacking disordered ices have been developed by various groups 
\citep[e.g.][and references therein]{malkin_2015, kuhs_2012}.
However, an in-depth analysis of the obtained diffraction patterns is 
not required for the purpose of this work and is not feasible 
using the moderate resolution data obtained.

Nevertheless, our molecular-scale neutron-scattering data are sufficient 
to characterize three distinct phase regimes for our icy particles
(Figures~\ref{fig:ice_phase} and~\ref{fig:ice_temp_change}). 
The temperature range, 
across which contributions from $I_{x}$ are seen, 
supports the interpretation as inter-domain ice (lacking long-range order) 
sandwiched between hexagonal and cubic domains,
as illustrated e.g.\ in Figures 2 and 3 of
\citet{hondoh_2015}.
These inter-domain amorphous structures; 
are not equivalent to, nor to be confused with, 
diffuse surface layers or vapor-deposited amorphous solid water (ASW).
We find that none of the phase-change temperatures matches the collisional 
temperature dependencies, which set in above $\approx210$~K.
Thus, we conclude that bulk crystalline ice-phase 
cannot influence collisional outcomes in our experiments and 
further crystallographic studies on a dedicated instrument 
are not required.

\subsection{Does the surface play a dominant role?}
\label{sec:surface_discussion}

Surface features could be connected 
to collision outcomes and particle aggregation in several ways. 
Both surface wetting and surface roughness might be 
expected to increase the stickiness of particles via friction effects. 
Molecular scale features ($\mathrm{\AA}$-scale) 
such as molecular orientation, mobility, or density variations 
on the surface might affect particle stickiness. 
They would affect the small angle scattering slope, but not the SSA.
Surface roughness on nm-scales would affect 
the small angle scattering slope and increase the observed SSA, 
particle sintering in the aggregation process would reduce the SSA 
with respect to that of smooth spherical particles. 

Based on the original Porod analysis of our small angle scattering data, 
we can exclude surface roughness, 
and indeed we find that even at the lowest temperatures 
the observed SSA is below that 
expected from the given size distributions of spherical particles 
by a factor of $\approx3$ for both mean particle sizes
(Figure~\ref{fig:SSA+thickness}), which could indicate particle sintering.
However, the SEM images (Figure~\ref{fig:ice_temp_change}(a$_2$)) 
reveal that at the lowest temperatures the particles are mostly, 
but not perfectly, spherical and in contact, 
slightly reducing the observable SSA, although no sintering is evident yet.

As the temperature is increased, we observe a gradual loss in SSA 
up to around 184~K (Figure~\ref{fig:SSA+thickness}),
which is explained by the onset of particle sintering, 
evident in the SEM images
(Figure~\ref{fig:ice_temp_change}(b$_2$)).
While such ice particles still show temperature independent outcomes 
in laboratory collision experiments
\citep{gundlach_2015}, 
the sintering of particles in close proximity, 
even at such low temperatures and pressures, 
corroborates earlier suggestions 
\citep{sirono_2017, sirono_2011b} 
that protoplanetary disk particles in contact will 
eventually merge together over long time scales, 
thereby forming conjoined objects.

Sintering becomes more pronounced with further increases in temperature
(Figure~\ref{fig:ice_temp_change}(c$_2$))
commensurate with the rapid SSA loss beyond 184~K
(Figure~\ref{fig:SSA+thickness}), 
and the particles shrink in diameter
(Figure~\ref{fig:ice_temp_change}(c) and (d)).
This suggests a loss of ice via sublimation, which agrees with 
our earlier finding that the normalized fractions of ice phases 
(Figure~\ref{fig:ice_phase})
sum to 1 only at the lowest temperature.  
This is also supported by the GudrunN processing of the original data: 
the total scattering at high $Q$ yields volume filling factors (not shown); 
these indicate which fraction of the probed volume is 
filled with sample material and 
also point to the amount of sample slightly decreasing with increasing $T$.

At the highest temperatures
(Figure~\ref{fig:ice_temp_change}(d$_2$)),
material is lost even from the sinter necks, 
commensurate with the decrease in the rate of SSA change,
observed via neutron scattering.

The SSA results come with one caveat;
the initial SSA is lower than could be explained by particle contact alone.
Possible reasons for this are:
baseline errors in the corrected neutron scattering data 
(attributable to potential H$_2$O ($\leq4~\%$) 
contamination in the samples, affecting the absolute $I(Q)$ calibration) 
could introduce a systematic error 
(up to a factor of 1.3) to the absolute SSA values, 
which therefore must be considered lower limits. 
The size distribution derived in earlier studies might be slightly altered 
by a different tube length between aspirator and liquid nitrogen Dewar 
or by size segregation effects during the filling process.
We assume that all of these affect the absolute SSA values to some degree 
and in combination explain the observed discrepancies.

While the absolute values of SSA might be affected 
by the above systematic deviations, the trends are not, 
and changes in the SSA are indicative of changes 
to the particle's surface structures, which might affect collision outcomes.

While both sintering and sublimation could affect particle stickiness, 
the temperature trends in the SSA and larger scale surface structures 
are not commensurate with those in collision experiments.
Thus, they cannot play a key role in determining collision outcomes.

We therefore return to the surface features on $\mathrm{\AA}$-scales: 
while the Porod analysis excluded surface roughness, it pointed to 
our particles having a diffuse interface at all temperatures observed.
The thickness $t$ of this interface starts to 
increase non-linearly with temperature above 206~K (Figure~\ref{fig:SSA+thickness}).
Our experiments were conducted over two neutron-beam periods. 
In each run, one sample of each particle diameter was investigated. 
Irrespective of particle diameter, the temperature trend in $t$ was 
similar between the two runs but more pronounced in the first run, 
leading to relatively large uncertainties on the average $t$, 
but not affecting the conclusions about the temperature ranges 
over which changes occur.

The increase in $t$ (Figure~\ref{fig:SSA+thickness}) 
indicates that the outermost water molecules become more mobile 
and more disordered than the bulk material. 
This phenomenon is well known from hail, ice and snow physics, 
where it is often referred to as surface pre-melting or quasi-liquid layers.
It is usually attributed to a reduction of the free surface energy of the ice 
by this re-organisation of molecules
\citep[][and references therein]{li_2007, dash_1995}, 
although the names are misleading, 
as the surface is not truly liquid in such cases. 

The observed temperature dependencies suggest 
that the $\mathrm{\AA}$-scale surface properties 
of icy particles are impacting on collision experiments. 
Below $\approx$~210~K collisional outcomes are temperature independent 
\citep{gundlach_2015, bridges_1996, hill_2015a}, 
and this equates to the regime where the thickness of the 
diffuse surface layer is invariant. 
Above $\approx$~210~K, the thickness of this layer increases 
(non-linearly with temperature) and temperature dependence is 
observed in collisional outcomes 
\citep{gundlach_2015}. 

Typical collisional studies would use H$_2$O samples, 
while our neutron scattering characterization required the use of D$_2$O. 
In general, 
the structural differences between the two materials are at a level of $<4$~\% 
in terms of the intramolecular bond length 
and even smaller on intermolecular scales
\citep{soper_2008}.
Isotope effects on molecular mobility are very complex, 
but overall the heavier D$_2$O molecules (20~amu) 
are less mobile than H$_2$O ones (18~amu).
For example at 298~K the diffusion coefficients are 
$\left(2.109\pm0.003\right)\times10^{-5}$~cm$^2$/s (D$_2$O) and 
$\left(2.272\pm0.003\right)\times10^{-5}$~cm$^2$/s (H$_2$O)
\citep{eisenberg_2005};
the triple point of D$_2$O is slightly higher (276.967~K)
\citep{marko_1989}
than that of H$_2$O (273.16~K).
Therefore, H$_2$O samples can be assumed to show 
slightly thicker diffuse interface layers than observed in this work, 
although the differences are likely to be at a level of a few percent only.

\subsection{Are the icy particles in laboratory collision experiments 
good analogs?}\label{sec:ice_phase_discussion}

Interstellar ices are either formed by 
vapor deposition of gas-phase water onto cooling dust grains 
\citep[e.g.][]{visser_2009},
or through the chemical vapor deposition of H and O atoms 
to eventually form water ice
\citep[e.g.][]{accolla_2013, cuppen_2010}.
Either way, the subsequent thermal or energetic processing of such ices 
results in crystallization of the material
\citep[e.g.][]{burke_2010, baragiola_2003},
so that both amorphous and crystalline ices have been detected 
in accretion disks around young stars
\citep{boogert_2015, terada_2012, schegerer_2010}. 

All ice collisional experiments to date 
\citep[e.g.][]{bridges_1984, deckers_2016}, 
including ours 
\citep[e.g.][]{gundlach_2015, hill_2015a}, 
must have been colliding crystalline ice particles, 
as no method has yet been reported to effectively produce 
amorphous particles suitable for collision experiments.

From this work we can now confirm that the trends 
observed in collision outcomes as a function of temperature 
cannot be attributed to changes in ice-phase-composition of crystalline ice.
Therefore, under suitable $P$-$T$ conditions all crystalline ice analogs, 
be they $I_c$, $I_h$, or a mixture of both, 
are well suited to replicate grain collisions in astrophysical environments 
where icy grains are dominated by crystalline icy material, 
e.g.\ heated regions of PPD or ring systems 
and planetary moons and atmospheres.
It remains an open question (but beyond the scope of this paper), 
whether collisional outcomes are the same for entirely amorphous icy particles or particles dominated by amorphous ice.

The key to icy particle collisional behavior, at least in laboratory experiments 
\citep[e.g.][]{gundlach_2015, heisselmann_2010, hill_2015a}, 
must be the diffuse surface layer, 
which increases the water-ice stickiness through surface pre-melting. 
This effect is known to be promoted by any type of irregularity 
at the surface, such as polycrystallinity, surface contact, 
or impurities. 
As our samples show stacking disorder and are granular, 
we would expect thicker diffuse surface layers in comparison to 
e.g.\ carefully prepared flat single crystal ice samples 
\citep{dosch_1995}, 
where it has previously been demonstrated that surface pre-melting 
can even occur at pressures as low as $P=6\times10^{-3}$~mbar, 
which is well within the pressure regime expected for a 
typical protoplanetary disk (Figure~\ref{fig:P_T_range}~(a)).

Figure~\ref{fig:P_T_range} summarizes the $P$-$T$ regimes 
for our, and other, planet-formation studies, 
putting them in context with the conditions 
typically found in proto-planetary disks. 
While our neutron scattering characterization was necessarily
done at a constant pressure (30 mbar He), given the 
experimental constraints we must assume that the critical temperature 
at which the thickness of the diffuse surface layer increases upon heating, 
will change with pressure and composition of the ambient gas. 
This is qualitatively depicted by the dark gray area 
(Figure~\ref{fig:P_T_range}), 
whose border connects our observation at 30~mbar to H$_2$O's triple point. 
The atmospheric pressure end of the curve is informed 
by earlier observations that H$_2$O molecules become mobile enough 
to restructure the surface and 
reduce the number of incompletely coordinated molecules at the boundary 
at temperatures as low as $60-120$~K 
\citep{devlin_2001}.

Most importantly, Figure~\ref{fig:P_T_range}~(a) shows, 
that all collision studies that resulted in a certain percentage of sticking 
were indeed carried out under $P$-$T$ conditions 
where a diffuse surface layer exists on the icy particles; 
and temperature dependence in the collisional outcomes 
is induced where the thickness of that layer starts to increase.
The one caveat to this is the work of 
\citet{musiolik_2016}, 
whose experiments were conducted under  $P$-$T$ conditions 
far from those where the diffuse surface layer dominates sticking outcomes. 
However, their experimental conditions relied on equilibria between gas, 
liquid, and solid water, which would result in 
dynamic exchange of water molecules at the particle surfaces, 
in a so-called dynamic liquid-like surface layer, 
as also reported in atmospheric studies of icy grains 
\citep[][and references therein]{li_2007, dash_1995}. 
Nevertheless, as Figure~\ref{fig:P_T_range}~(a) shows, care 
should be taken when considering the outcomes of ice collision studies, 
as the influence of the $P$-$T$ conditions on the surface structure 
and behavior of the particles seems, from this work, to be 
at least as important as the velocity or size of the particles. 
The open question remains to what extend the diffuse surface layer impacts 
on collision outcomes under the $P$-$T$ conditions in a protoplanetary disk.


\section{Conclusions}\label{sec:conclusions}

We have characterized the ice particle analogues 
used in our laboratory collision experiments 
\citep{gundlach_2015},
exploiting neutron scattering and cryo-SEM to determine whether 
they are good analogs for protoplanetary disk environments, 
whether their ice phase affects the collisional outcome, 
and whether their surface structure plays a dominant role in collisions.

Our analysis shows that 
neither changes in specific surface area nor in crystalline ice phase tie-in 
with previously observed temperature dependencies of collisional outcomes.
The key to these temperature effects must be 
the increasing thickness of the diffuse surface layer, 
which at 30~mbar pressure is present across all 
investigated temperatures ($103-247$~K), 
but starts to increase in thickness above $\approx210$~K, 
matching the observed onset of temperature dependent collision outcomes.

Ideally, experiments would  always be performed at 
$P$-$T$ conditions that are expected for proto-planetary disks. 
Where that is prevented by the experimental procedures, 
care should be taken to avoid $T$-dependent outcomes 
induced by the diffuse surface layer. 
Therefore, at pressures of a few mbar, 
collision experiments should be performed below 210~K, 
as changing surface structural properties of the ice 
will otherwise affect the collision outcomes.

While the typical production methods for crystalline ice analogs 
do not impact on the collision outcomes, the collision environment does.
Therefore, in laboratory studies of icy particle collisions 
the parameter space, particularly with reference to $P$ and $T$, 
must be clearly defined and controlled.
In collisional studies to date significant parameter space 
w.r.t pressure, temperature, size, and velocity is yet unexplored 
(Figure~\ref{fig:P_T_range})
and particle sticking has so far only been observed 
at pressures higher than expected in protoplanetary disks.
Unlike Silicon-based dust particles, bulk and surface structures of ice are 
influenced by the surrounding $P$-$T$ conditions. 
Nevertheless planet formation models currently prioritize parameters 
such as particle velocity and size. 
Further investigation of the $P$-$T$ range at which 
diffuse surface layers affect collision outcomes is clearly warranted, 
to enable such findings to be incorporated into future models.


\acknowledgments

We kindly acknowledge helpful discussions with 
Richard Heenan (ISIS Facility).
Experiments at the ISIS Pulsed Neutron and Muon Source 
were supported by a beamtime allocation on the 
near and intermediate range order diffractometer NIMROD from the 
Science and Technology Facilities Council, RB1520425.
SG and HJF acknowledge support from The Open University. 
Astrochemistry at the Open University is supported by STFC under 
grant agreements No. ST/M007790/1, ST/M003051/1, ST/N006488/1, 
ST/N005775/1 and ST/L000776/1 as well as 
Royal Society International Exchange Award (IE/14/3). 
SG and HJF both gratefully acknowledge travel support associated with 
this project from the EU COST Action CM1401 Our Astrochemical History, 
including STSM - 100615-061929 linked with this work.
BG, JR and JB acknowledge support from TU Braunschweig and DLR 
under grant 50WM1536.

%



\software{GudrunN \citep{soper_2011, soper_2013},  
              }




%



\bibliography{Icy_Particles}



\end{document}